\documentclass[pra,twocolumn,showpacs,superscriptaddress,10pt]{revtex4-1}
\usepackage[colorlinks,citecolor=blue,linkcolor=red,dvipdfm]{hyperref}
\usepackage{amsmath,amssymb,graphicx}
\usepackage{times}

\begin{document}

\title{A purely Kerr nonlinear model admitting flat-top solitons}
\author{Liangwei Zeng$^{1,2}$, Jianhua Zeng$^{1,2,6}$, Yaroslav V. Kartashov$^{3}$, and
Boris A. Malomed$^{4,5}$}
\address{
$^{1}$State Key Laboratory of Transient Optics and Photonics, Xi'an Institute of Optics and Precision Mechanics of CAS, Xi'an 710119, China\\
$^{2}$University of Chinese Academy of Sciences, Beijing 100084, China\\
$^{3}$Institute of Spectroscopy, Russian Academy of Sciences, Troitsk, Moscow, 108840, Russia\\
$^{4}$Department of Physical Electronics, School of Electrical Engineering, Faculty of Engineering, Tel Aviv University, Tel Aviv 69978, Israel\\
$^{5}$ITMO University, St. Petersburg 197101, Russia\\
$^{6}$Corresponding author: \underline{zengjh@opt.ac.cn}}

\begin{abstract}
We elaborate one- and two-dimensional (1D and 2D) models of media with self-repulsive cubic nonlinearity, whose local strength is subject to spatial modulation that admits the existence of flat-top solitons of various types, including fundamental ones, 1D multipoles, and 2D vortices. Previously, solitons of this type were only produced by models with competing nonlinearities. The present setting may be implemented in optics and Bose-Einstein condensates. The 1D version gives rise to an exact analytical solution for stable flat-top solitons, and generic families may be predicted by means of the Thomas-Fermi approximation. Stability of the obtained flat-top solitons is analyzed by means of linear-stability analysis and direct simulations. Fundamental solitons and 1D multipoles with $k=1$ and $2$ nodes, as well as vortices with winding number $m=1$, are completely stable. For multipoles with $k\geq 3$ and vortices with $m\geq 2$, alternating stripes of stability and instability are identified in their parameter spaces.
\end{abstract}
\maketitle

The formation of bright spatial solitons in one-dimensional (1D) uniform media is a commonly
known result of the balance between the diffraction and self-focusing
nonlinearity \cite{FPC}; however, making them stable in higher-dimensional media is a challenging issue owning to the presence of the wave collapse, which
can, in an usual way, be suppressed with an aid of linear periodic
potentials \cite{FPC, NL}. The situation may be different in physical settings with inhomogeneous strength of the local nonlinearity \cite{NL}. In optics, such settings may be engineered by means of properly designed photonic-crystal
structures, with voids filled by solid \cite{Russell} or liquid \cite%
{phot-cryst1,phot-cryst2,phot-cryst3} materials with different values of the
Kerr coefficient. Alternatively, one can use nonuniform distributions of
nonlinearity-enhancing dopants \cite{dopant,Gaetano}. In Bose-Einstein
condensates (BECs), similar nonlinearity landscapes can be created by means
of the Feshbach resonance (FR)\ locally controlled by spatially nonuniform
optical \cite{nonuniform-Feshbach1, nonuniform-Feshbach2,
nonuniform-Feshbach3} or magnetic \cite{Feshbach-magnetic1,
Feshbach-magnetic2} fields. In particular, it was predicted that
self-defocusing nonlinearity, whose local strength grows from the center to
periphery in the $D$-dimensional space, with radial coordinate $r$, at any
rate faster than $r^{D}$, can support a great variety of robust self-trapped
modes, including 1D fundamental, dipole and multipole solitons, 2D solitary
vortices with arbitrarily high topological charge $m$ \cite{Olga,Olgaalg}, and
sophisticated 3D modes, such as soliton gyroscopes \cite{Defo9} and
skyrmions \cite{Defo8}. A characteristic feature of these localized modes is
\textit{nonlinearizability} of the underlying equations for their decaying
tails, on the contrary to the usual bright solitons maintained by uniform
self-focusing, whose exponential tails are produced by the corresponding linearized
equations. The exploration of different kinds of bright solitons supported by this scheme has
currently been extended to a variety of other physical settings \cite{Defo5}-\cite{Defo14}.

The objective of this Letter is to demonstrate another natural setting,
which gives rise to families of 1D and 2D \textit{flat-top} solitons. This
is a known variety of self-trapped modes, with a potential for applications
\cite{Porras}, which are usually supported by systems with competing
focusing and defocusing nonlinearities \cite{New}, such as cubic-quintic
\cite{Bulgaria,Humberto,Humberto2,nine-authors}, cubic-quartic \cite%
{Petrov,we}, and quadratic-cubic \cite{chi2,Grisha1,Grisha2} combinations,
as well as by cubic terms with an additional logarithmic factor \cite%
{Grisha1,NJP} (the two former types of the nonlinearity were recently
realized experimentally in optics \cite{Cid} and as \textquotedblleft
quantum droplets" BEC \cite{droplets1,droplets2}, respectively). However,
stable flat-top solitons were not previously found in physical models with the
cubic-only nonlinearity, while here we demonstrate that this is possible, in
1D and 2D geometries alike, for various soliton species (fundamental,
multipole, vortical), if the coefficient of the cubic defocusing is
subjected to an appropriate spatial modulation. In addition to the systematically produced numerical results, we also obtain particular exact stable solutions for the 1D fundamental flat-top solitons, and develop the Thomas-Fermi (TF) approximation for generic soliton states and 2D vortices. Stability of all the solutions is investigated via the linear-stability analysis and numerical simulations.

The basic model is introduced as the scaled Schr\"{o}dinger equation
governing the evolution of the dimensionless complex amplitude of a light
beam propagating in a cubic nonlinear medium, or the mean-field wave
function in a Bose-Einstein condensate (BEC), $\psi \left( \mathbf{r}
,t\right) $:
\begin{equation}
i\psi _{t}=-(1/2)\nabla ^{2}\psi +\sigma (r)\left\vert \psi \right\vert
^{2}\psi ,  \label{eq:NLSE}
\end{equation}
which is written in the 2D form, with $\mathbf{r}=\left( x,y\right) $. Here $
t$ is the evolution variable, representing time in the BEC version of the
model, or the propagation distance in optics. The axially symmetric
nonlinearity-modulation profile, $\sigma (r)>0$, is chosen in the form which
readily helps to create flat-top modes of radius $r_{0}$,
\begin{equation}
\sigma (r)=\left\{
\begin{array}{c}
\sigma _{\mathrm{int}},r\leq r_{0}, \\
\sigma _{\mathrm{ext}}\exp [\alpha (r+\alpha ^{-1/2}-r_{0})^{2}-1],r>r_{0},
\end{array}
\right.   \label{eq:gprofile}
\end{equation}
with constants $\sigma _{\mathrm{int,ext}}>0$ and $\alpha >0$. Note that Eq.
(\ref{eq:gprofile}) implies that $\sigma (r=r_{0})=\sigma _{\mathrm{ext}}$.
The 1D version of the model corresponds to Eq. (\ref{eq:NLSE}) with single
coordinate $x$, while $r$ and $r_{0}$ are replaced by $|x|$ and $x_{0}>0$ in
Eq. (\ref{eq:gprofile}). This profile can be created by means of the
above-mentioned methods -- for instance, by the application of an
FR-controlling magnetic-field profile to the BEC layer, with a constant
detuned value at $r<r_{0}$, and one approaching the exact resonant value at $
r>r_{0}$. It is relevant to mention that similar cylindrical optical-box
potentials are used in current BEC\ experiments \cite{Navon}.

Wave functions of stationary states with real chemical potential $\mu $
(alias propagation constant $-\mu $ in terms of optics) and integer
vorticity $m$ are sought for, in ploar coordinates $\left( r,\theta \right) $
, as $\psi \left( \mathbf{r},t\right) =w(r)\exp \left( im\theta -i\mu
t\right) $, with real $w$ determined by the equation
\begin{equation}
\mu w=-\left( 1/2\right) (w^{\prime \prime }+r^{-1}w^{\prime
}-m^{2}r^{-2}w)+\sigma (r)w^{3}  \label{w}
\end{equation}
in 2D, or its counterpart in 1D. First, the 1D version of the model admits
an exact solution, with $\mu =3\alpha /2$, $w\left( x\leq \left\vert
x_{0}\right\vert \right) =\sqrt{3\alpha /2\sigma _{\mathrm{ext}}}$, and

\begin{gather}
w(|x|>x_{0})=(\alpha /\sqrt{2\sigma _{\mathrm{ext}}})(|x|+\alpha
^{-1/2}-x_{0})  \nonumber \\
\times \exp (-(1/2)[\alpha (|x|+\alpha ^{-1/2}-x_{0})^{2}-1]).
\label{exact1d}
\end{gather}
Note that Eq. (\ref{exact1d}) gives $dw/dx\left( x=x_{0}\right) =0$, which
is necessary for the continuity at $x=x_{0}$, while the remaining continuity
condition for $w(x)$ imposes a relation on constants of modulation profile (
\ref{eq:gprofile}): $\sigma _{\mathrm{int}}=3\sigma _{\mathrm{ext}}$. This
profile is continuous in the limit case of $x_{0}=0$, for which exact
solution (\ref{exact1d}) remains valid. Numerical solutions are displayed
below for the continuous modulation profile with $\sigma _{\mathrm{int}
}=\sigma _{\mathrm{ext}}$.

An analytical approximation for generic soliton shapes can be obtained in
the Thomas-Fermi (TF) approximation, which neglects derivatives in Eq. (\ref
{w}) \cite{Fetter}:
\begin{equation}
w_{\mathrm{TF}}^{2}=\left[ \sigma (r)\right] ^{-1}[\mu -m^{2}/(2r^{2})]
\label{TF}
\end{equation}
at $r^{2}>m^{2}/2\mu $, and $w_{\mathrm{TF}}^{2}=0$ at $r^{2}<m^{2}/2\mu $.
This approximation makes it possible to predict the dependence of the norm
of the soliton family on the chemical potential,
\begin{gather}
N_{\mathrm{2D}}=2\pi \int_{0}^{\infty }w^{2}(r)dr\approx \pi \sigma _{
\mathrm{int}}^{-1}\times   \nonumber \\
\lbrack \mu r_{0}(r_{0}+\alpha ^{-1/2})-(m^{2}/2)\ln (2e\mu r_{0}^{2}/m^{2})]
\label{NTF}
\end{gather}
(here $e$ is the base of the natural logarithm), which is valid at $\mu
>m^{2}/\left( 2r_{0}^{2}\right) $.

In the numerical form, stationary profiles of both 1D and 2D modes were
found by means of the Newton's method applied to Eq. (\ref{w}). The
subsequent stability analysis was based on the usual ansatz, $\psi
=[w(r)+p_{+}(r)\mathrm{exp}(in\theta +\lambda t)+p_{-}^{\ast }(r)\mathrm{exp}
(-in\theta +\lambda ^{\ast }t)]\mathrm{exp}(im\theta -i\mu t)$, where $
p_{\pm }(r)$ represent perturbation eigenmodes with eigenvalue $\lambda $, $
\ast $ stands for the complex conjugate, and $n$ is an integer azimuthal
index. Then, the eigenvalue problem amounts to the solution of linear
equations,

\begin{gather}
i\lambda p_{\pm }=\mp \left( 1/2\right) [ p_{+}^{\prime \prime
}+r^{-1}p_{\pm }^{\prime }-(m\pm n)^{2}r^{-2}p_{\pm }] p  \nonumber \\
\mp \mu p_{\pm }+\sigma w^{2}(2p_{\pm }+p_{\mp }),  \label{pert}
\end{gather}
or their 1D counterparts. In particular, the exact solution given by Eq. (
\ref{exact1d}) is found to be always stable.

\begin{figure}[tbp]
\begin{center}
\includegraphics[width=1\columnwidth]{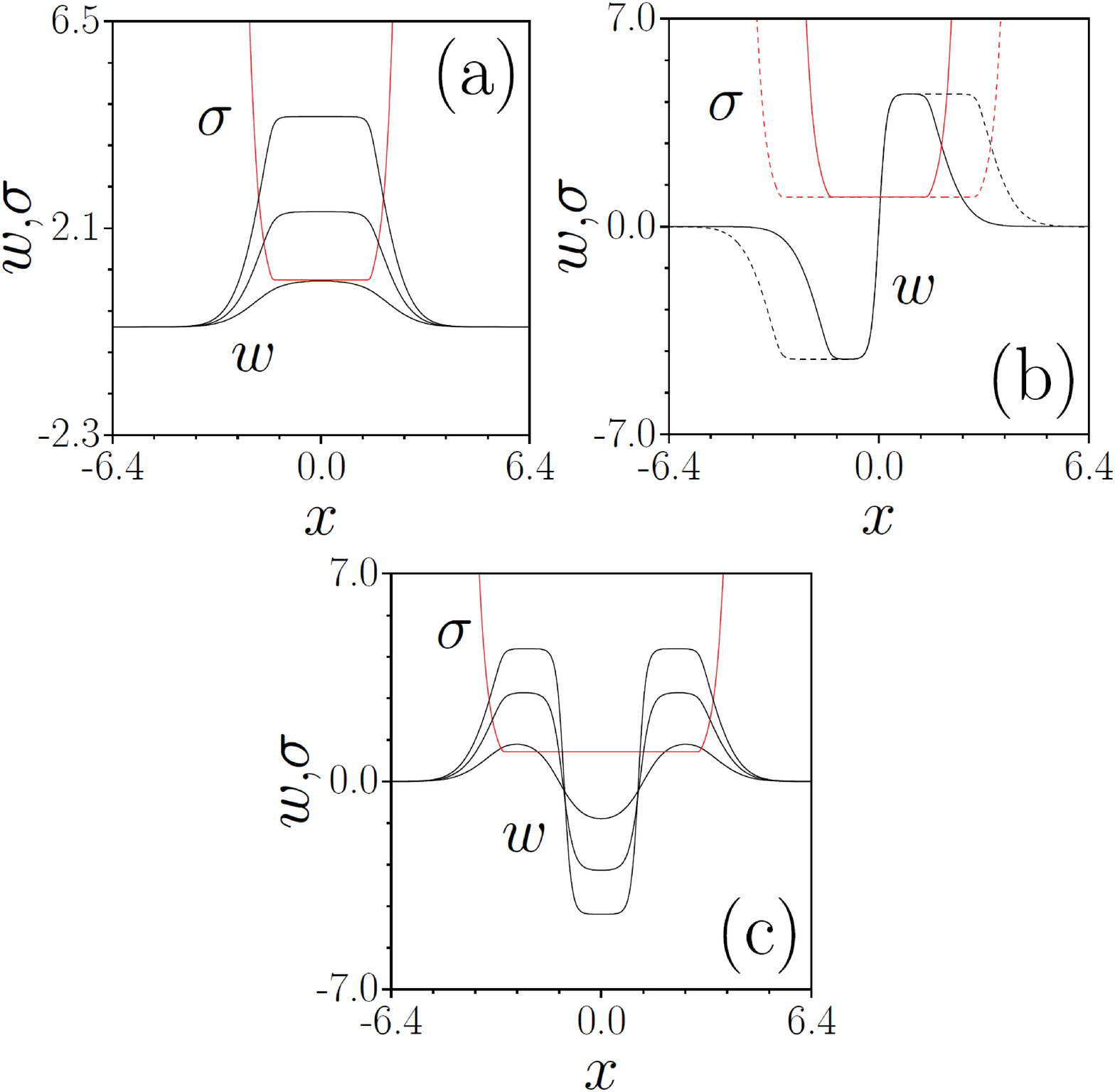}
\end{center}
\caption{(a) Fundamental flat-top ($k=0$) solitons with $\protect\mu =1,6,20$
at $x_{0}=1.5$. (b) Flat-top dipole ($k=1$) solitons with $\protect\mu =20$
at $x_{0}=1.5$ and $3$ (solid and dashed lines, respectively) (c) The
transition of tripole ($k=2$) modes into the flat-top ones with increasing $
\protect\mu =2,9,$ and $20$ at $x_{0}=3$. Here and in Fig. \protect\ref{fig4}
, red lines represent the nonlinearity-modulation profiles given by Eq. (
\protect\ref{eq:gprofile}) with $\protect\alpha =\protect\sigma _{\mathrm{
ext,int}}=1$ (these values are used throughout the work). All modes with $
k=0,1,$ and $2$ are stable.}
\label{fig1}
\end{figure}

\begin{figure}[tbp]
\begin{center}
\includegraphics[width=1\columnwidth]{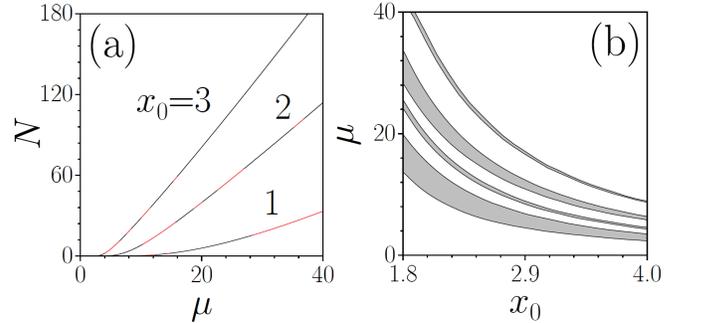}
\end{center}
\caption{(a) The norm versus $\protect\mu $ for 1D multipoles with $k=6$ at
different values of $x_{0}$. Here and in Fig. \protect\ref{fig5}(a), stable
and unstable segments are distinguished by the black and red colors,
respectively. (b) Stability (white) and instability (gray) domains for the
modes with $k=6$ in the $(x_{0},\protect\mu )$ plane.}
\label{fig2}
\end{figure}

\begin{figure}[tbp]
\begin{center}
\includegraphics[width=1\columnwidth]{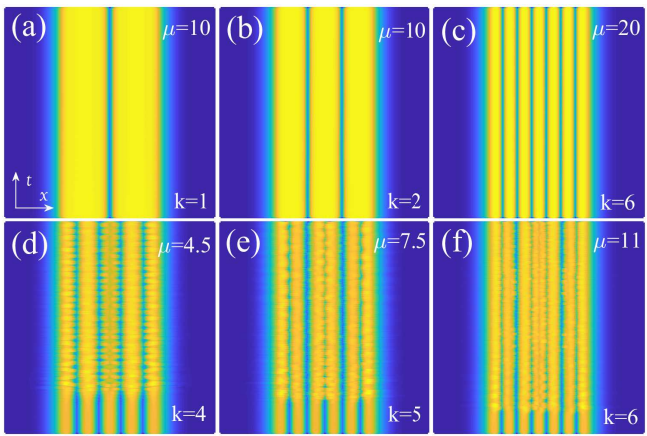}
\end{center}
\caption{Stable evolution of perturbed 1D solitons, for $x_{0}=3$: (a) a
dipole ($k=1$) with $\protect\mu =10$; (b) a tripole ($k=2$) with $\protect
\mu =10$; (c) a multipole ($k=6$) with $\protect\mu =20$. The evolution of
unstable multipoles, also for $x_{0}=3$: (d) a $k=4$ with $\protect\mu =4.5$
; (e) $k=5$ with $\protect\mu =7.5$; (f) $k=6$ with $\protect\mu =11$. The
evolution range is $0<t<1000$.}
\label{fig3}
\end{figure}

\begin{figure}[tbp]
\begin{center}
\includegraphics[width=1\columnwidth]{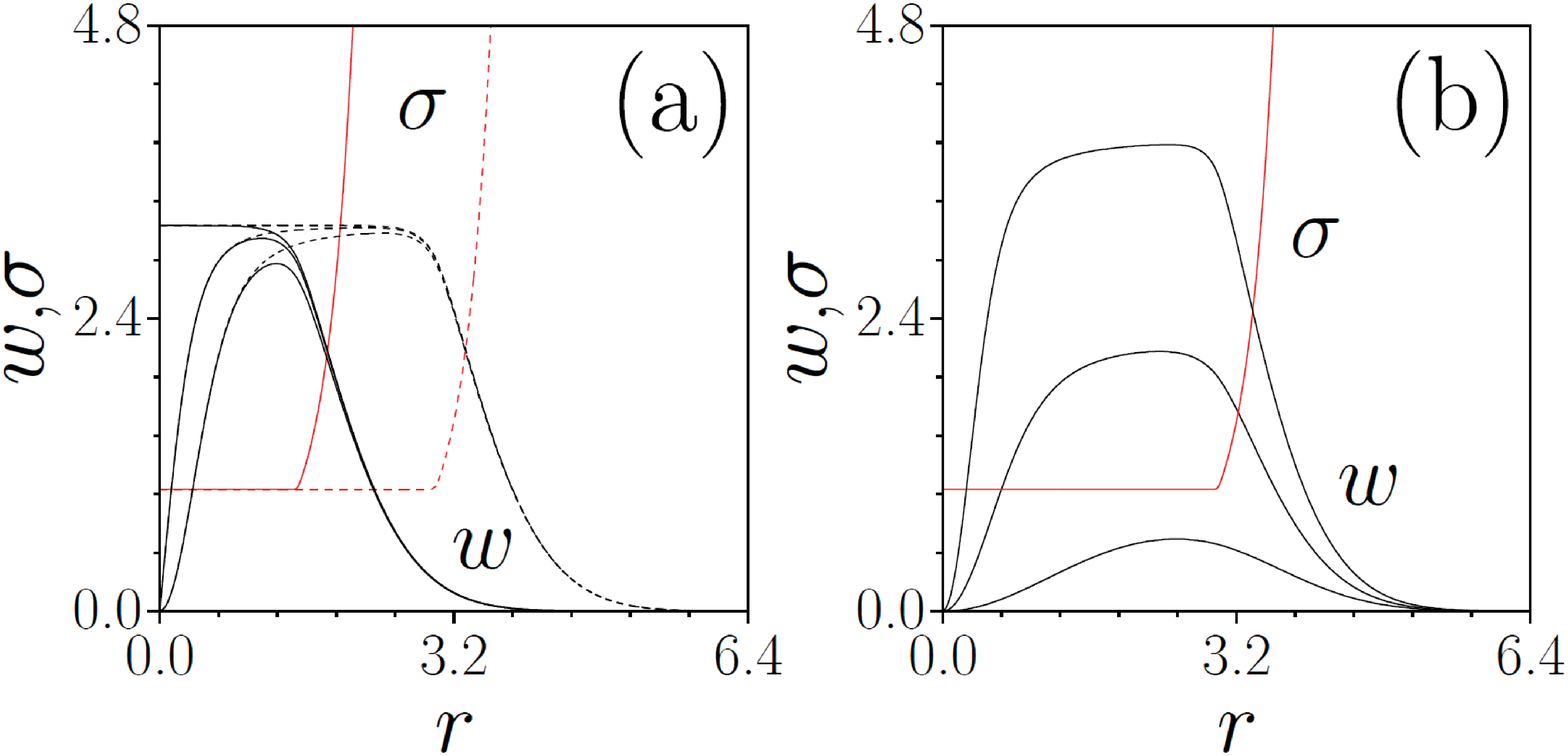}
\end{center}
\caption{(a) Profiles of 2D solitons with vorticities $m=0,1,2$ and $\protect
\mu =10$ at $r_{0}=1.5$ and $3$ (solid and dashed lines, respectively). (b)
The transition of vortices with $m=2$ into the flat-top shape with
increasing $\protect\mu =1,5,15$ at $r_{0}=3$.}
\label{fig4}
\end{figure}

Typical profiles of 1D flat-top solitons with the number of nodes $k=0$, $1$
, and $2$ (fundamental, dipole, and tripole solitons, respectively) at
different values of $\mu $ are displayed in Fig. \ref{fig1}, which clearly
shows that the solitons' shape gets flatter with the increase of $\mu $. The functional form of soliton changes considerably as $\mu$ increases. For instance, solitons with large number of nodes $k$ at small values of $\mu$ resemble trigonometric functions, while at large $\mu$ values they can be viewed as complexes of several well-localized dark solitons. Families of all solitons with $k\leq 2$ are completely stable (at least, up
to $\mu =40$), while instability domains appear at $k\geq 3$. To illustrate
this feature, Fig. \ref{fig2}(a) represents soliton families for $k=6$ by
showing their norm vs. $\mu $ at several values of $x_{0}$. It is seen that
even this high value of $k$, corresponding to \textquotedblleft hashed"
flat-top patterns, admits large stability segments, whose share increases
with the growth of width $x_{0}$ of the modulation profile. Note also that
all the $N(\mu )$ curves satisfy the \textquotedblleft
anti-Vakhitov-Kolokolov" criterion, $d\mu /dN>0$, which is a necessary
condition for the stability of solitons in models with self-repulsive
nonlinearities \cite{antiVK}. The alternating stability and instability
domains for $k=6$ are charted in the ($x_{0}$,$\mu $) plane in Fig. \ref
{fig2}(b). Figure \ref{fig2} shows that flat-top states are generally more
stable than their more localized counterparts with the same number of nodes $
k$.

Typical examples of the evolution of stable 1D flat-top solitons are
displayed in Figs. \ref{fig3}(a)-(c), while evolution of their unstable
counterparts is shown in Figs. \ref{fig3}(d-f). Unstable multipoles
spontaneously develop oscillations, keeping the number of nodes.

The profiles of 2D solitons with vorticities $m=0$ (fundamental solitons), $
m=1$, and $m=2$ for different values of $r_{0}$ are displayed in Fig. \ref
{fig4}(a), which shows that the width of the flat-top solitons increases
with $r_{0}$, similar to their 1D counterparts. Further, Fig. \ref{fig4}(b)
displays profiles of the vortices with $m=2$ and different values of $\mu $,
demonstrating that 2D solitons also get flatter with the increase of $\mu $.
\begin{figure}[tbp]
\begin{center}
\includegraphics[width=1\columnwidth]{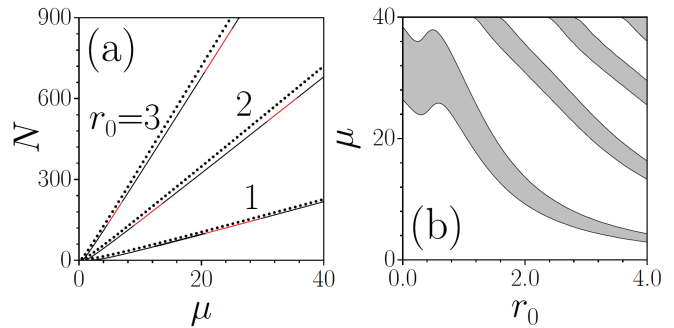}
\end{center}
\caption{(a) The norm of 2D double-vortex solitons ($m=2$) versus $\protect
\mu $ at different values of width $r_{0}$ of the nonlinearity-modulation
profile. Dotted lines display the prediction of the TF approximation, as per
Eq. (\protect\ref{NTF}). (b) Stability (white) and instablility (gray)
domains for the solitons with $m=2$ in the $(r_{0},\protect\mu )$ plane.}
\label{fig5}
\end{figure}
\begin{figure}[tbp]
\begin{center}
\includegraphics[width=1\columnwidth]{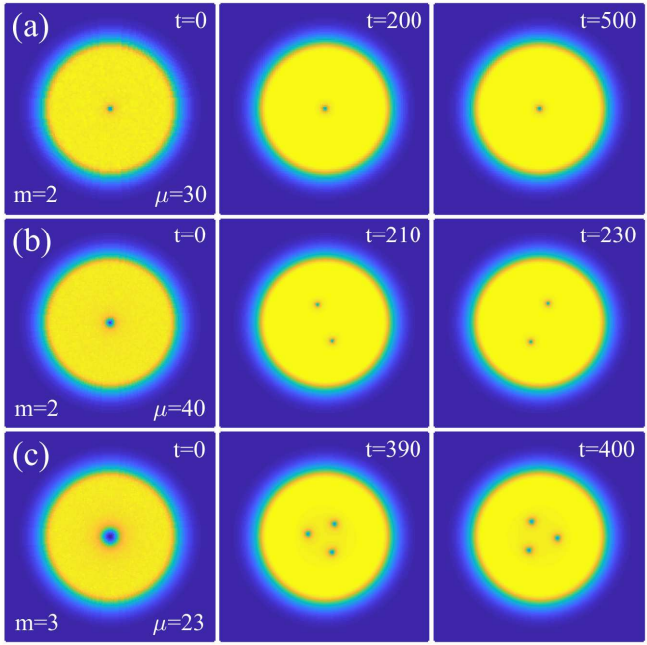}
\end{center}
\caption{The evolution of perturbed 2D vortex solitons at $r_{0}=4$: (a) a
stable soliton with $m=2$, $\protect\mu =30$; (b) an unstable soliton with $
m=2$, $\protect\mu =40$; (c) an unstable one with $m=3$, $\protect\mu =23$.}
\label{fig6}
\end{figure}

Typical dependencies $N(\mu )$ for the vortex families with $m=2$ are
displayed in Fig. \ref{fig5}(a), featuring a nearly linear form for all
values of $r_{0}$, and considerable growth of $N$ with the increase of $r_{0}
$. These features are well predicted by the TF approximation, as seen in the
figure. The 2D modes with $m=0$ and $1$ are completely stable, at least up
to $\mu =40$, while the vortices with $m\geq 2$ demonstrate alternation of
stability and instability domains in the ($r_{0}$, $\mu $) plane in Fig. \ref
{fig5}(b). Finally, the evolution of the flat-top 2D vortices is displayed
in Fig. \ref{fig6}, demonstrating that those with $m=2$ or $3$, which are
unstable, split into persistently rotating pairs or triplets of unitary
vortices.

In conclusion, we have demonstrated that families of stable flat-top
solitons, including 1D multipoles and 2D vortices, can be created in media
with cubic self-repulsive nonlinearity whose local strength is subject to an
appropriate spatial modulation. To our knowledge, this model predicts the
first stable flat-top solitons realized with the cubic-only nonlinearity, in contrast to previous results which demonstrated such modes solely in
systems with competing attractive and repulsive nonlinearities, thus providing an alternative way to create and stabilize
the flat-top modes with free intrinsic parameters. We have
checked the stability of all the obtained flat-top solitons
by means of the linear-stability analysis and direct
simulations.
Both the 1D and 2D solitons become flatter with the increase of their
chemical potential. 1D multipoles with $k=0,1,$ and $2$ nodes, as well as 2D
solitons with vorticities $m=0$ and $1$ are completely stable, while
higher-order modes, with $k\geq 3$ and $m\geq 2$, respectively, feature
alternating stability and instability domains. Such self-trapped modes can
be created in BEC and optics by means of available experimental techniques.


The work of LZ and JZ was supported by the NSFC (Nos. 61690224, 61690222),
and by the Youth Innovation Promotion Association of the Chinese Academy of
Sciences (No. 2016357). The work of B.A.M. was partly supported by the
Israel Science Foundation through grant No. 1287/17.

\bigskip



\end{document}